\documentclass[sigconf]{acmart}
    
\usepackage{nicefrac}
\usepackage{siunitx}
\usepackage{array,framed}
\usepackage{booktabs}
\usepackage{
  color,
  float,
  epsfig,
  wrapfig,
  graphics,
  graphicx,
  subcaption
}
\usepackage{textcomp}
\usepackage{setspace}
\usepackage{latexsym,fancyhdr,url}
\usepackage{enumerate}
\usepackage{algorithm2e}
\usepackage{algpseudocode}
\usepackage{graphics}
\usepackage{xparse} 
\usepackage{xspace}
\usepackage{multirow}
\usepackage{csvsimple}
\usepackage{balance}

\usepackage{
  tikz,
  pgfplots,
  pgfplotstable
}
\usepackage{hyperref}

\usetikzlibrary{
  shapes.geometric,
  arrows,
  external,
  pgfplots.groupplots,
  matrix
}

\pgfplotsset{compat=1.9}

\usepackage{mathtools,}

\DeclareMathAlphabet{\mathcal}{OMS}{cmsy}{m}{n}

 \usepackage{hyperref} 
\AtBeginDocument{%
  \providecommand\BibTeX{{%
    \normalfont B\kern-0.5em{\scshape i\kern-0.25em b}\kern-0.8em\TeX}}}
 
\usepackage{booktabs}
\usepackage{multirow}
\usepackage{soul}
\usepackage{pifont}
\newcommand{\cmark}{\ding{51}}%
\usepackage{tikz}
\usepackage{amsmath} 

\usepackage[shortlabels]{enumitem}
\setitemize{noitemsep,topsep=0pt,parsep=0pt,partopsep=0pt}
\setenumerate{noitemsep,topsep=0pt,parsep=0pt,partopsep=0pt}

\setcopyright{acmcopyright}
\copyrightyear{2023}
\acmYear{2023}
\acmDOI{XXXXXXX.XXXXXXX}
\acmConference[CCS '23]{2023 ACM SIGSAC Conference on Computer and Communications Security}{November 26-30, 2023}{Copenhagen, Denmark}
\acmPrice{15.00}
\acmISBN{978-1-4503-XXXX-X/18/06}

\begin{document}

\title[Comprehension from Chaos: Towards Informed Consent for Private Computation]{Comprehension from Chaos:\\ Towards Informed Consent for Private Computation}

\author{Bailey Kacsmar}
\affiliation{%
  \institution{University of Alberta}
  \country{Canada}
}

\author{Vasisht Duddu}
\affiliation{%
  \institution{University of Waterloo}
  \country{Canada}
}

\author{Kyle Tilbury}
\affiliation{%
  \institution{University of Waterloo}
  \country{Canada}
}

\author{Blase Ur}
\affiliation{%
  \institution{University of Chicago}
  \country{USA}
}

\author{Florian Kerschbaum}
\affiliation{%
  \institution{University of Waterloo}
  \country{Canada}
}

\begin{abstract}
Private computation, which includes techniques like multi-party computation and private query execution, holds great promise for enabling organizations to analyze data they and their partners hold while maintaining data subjects' privacy. Despite recent interest in communicating about differential privacy, end users' perspectives on private computation have not previously been studied. To fill this gap, we conducted 22 semi-structured interviews investigating users' understanding of, and expectations for, private computation over data about them. Interviews centered on four concrete data-analysis scenarios (e.g., ad conversion analysis), each with a variant that did not use private computation and another that did (private set intersection, multi-party computation, and privacy preserving query procedures). While participants struggled with abstract definitions of private computation, they found the concrete scenarios enlightening and plausible even though we did not explain the complex cryptographic underpinnings. Private computation increased participants' acceptance of data sharing, but not unconditionally; the purpose of data sharing and analysis was the primary driver of their attitudes. Through collective activities, participants emphasized the importance of detailing the purpose of a computation and clarifying that inputs to private computation are not shared across organizations when describing private computation to end users.
\end{abstract}



\maketitle

\section{Introduction}\label{sec:intro}

As data access and collection have grown, so have companies' attempts to leverage that data, with regulations trailing far behind. 
Collaborations between companies increasingly involve data sharing and disclosure. For example, Mastercard raised privacy concerns when it sold transaction data to Google to track whether Google ran digital ads that led to a sale at a physical store (ad conversion)~\cite{bergen2018google}.

Within such modern data-sharing practices, a \textbf{data subject} is an entity whose data is present in the data set, while a \textbf{data controller} is an entity holding a data set. 
Data controllers who are not themselves the data subject may have different privacy expectations or requirements compared to when data subjects themselves directly make data-sharing decisions. 
The data subject may not have understood their data could even be shared or sold~\cite{voigt2017eu, linden2018privacy,fabian2017large, rader2014awareness}.


Private computation, encompassing complex cryptographic techniques like private set intersection (\textbf{PSI})~\cite{chen2018labeled, pinkas2018Euro} and multi-party computation (\textbf{MPC})~\cite{gmw,yao}, allows companies to analyze data while maintaining data subjects' privacy in many cases. Private computation is especially valuable for cases where the data is sensitive (e.g., health or financial data)~\cite{truex2019}, among mutually suspicious entities~\cite{cristofaro2010practical,camenisch2009private}, or when there are less open trust boundaries~\cite{truex2019}. 

For example, PSI refers to a computation where two or more parties who each hold a private data set wish to collectively compute the intersection of their sets. The intersection can then be shared with one or more of the participating parties. 
For example, two companies could determine which users they have in common without disclosing the identities of the users not in common. PSI, as with many other private computations, can be implemented using homomorphic encryption 
or various other techniques. 
The privacy guarantees provided follow from the specific mechanisms used and are based on 
statistical assumptions or computational hardness. 

While private computation is often substantially more computationally expensive and complex than its non-private analogue, there is an assumption that it is in some way \emph{better}. For instance, it is presumed to be better for privacy that when PSI is used, data is only shared about clients the organizations have in common. 
To date, the degree to which users perceive private computation as better, or even feasible and plausible, has remained an open question. 

Furthermore, despite a flurry of recent work investigating users' expectations of differential privacy~\cite{xiong2020towards,kuhtreiber2021usable,kuhtreiber2022replication,Bullek2017} and attempting to improve communication about differential privacy~\cite{Karegar2022,Cummings2021,nanayakkara2022visualizing,CORMAN2022104859,franzen2022private}, 
users' attitudes about---and expectations for---the broader range of techniques subsumed under private computation has remained open. The only user-centered work on private computation~\cite{agrawal2021exploring,smart2022understanding} has investigated usability from an expert's, rather than an end user's, perspective. While explanations of differential privacy for end users often try to convey the intuition of adding noise or randomness to data or to a computation, the underlying mathematics of other private computation techniques lack an intuitive analogue, yet the guarantees and benefits are arguably more straightforward.

To recap, when an organization considers deploying private computation, two key attributes must be addressed. First, what privacy guarantees can actually be made to data subjects? Second, are those guarantees meaningful to the data subjects whose privacy they aim to protect? 
In this work, we investigate the second question through 22 semi-structured interviews. 
Without knowing what data subjects understand and expect from private computation, one cannot develop tools that empower them to make informed choices. 
Thus, in this paper we ask and answer the following research questions (\textbf{RQs}): 
\begin{itemize}
    \item \textbf{RQ~1}: What do data subjects understand about private computation, and how can specific examples facilitate their understanding of the concept? \emph{See Section~\ref{subsect:learn}}.
    \item \textbf{RQ~2}: How is a data subjects' willingness to share their data impacted when informed of private computation's protections and guarantees? \emph{See Sections~\ref{sect:initialperceptions}--\ref{subsect:conditions}}. 
    \item \textbf{RQ~3}:  How do data subjects perceive private computation's risks (inference attacks and beyond)? \emph{See Sections~\ref{subsect:risksImplications}--\ref{sect:inferenceattacks}.}
    \item \textbf{RQ~4}: How are perceptions of companies influenced by their use of private computation? \emph{See Section~\ref{subsect:responsibilities}.}
\end{itemize}

In brief, we found the following implications for private computation in practice.
First, data subjects are able to evaluate and understand the implications of private computation over their own data. Thus, neglecting to inform them of such practices is denying them autonomy over their own data. Second, while participants have an appreciation for the protections private computation can produce, they do not find these protections sufficient to overcome the need for both consent and transparency. That is, key details factor into participants' evaluation of acceptability (Section\ref{subsect:conditions}), and companies should communicate them.
Third, participants are aware of unique high-risk threat models against which private computation cannot guarantee protection (Section~\ref{subsect:risksImplications}). Thus, failing to communicate the implications of common private computation practices can create unintended risks for users and companies.

\section{Background}\label{sect:background}

Private computation is the suite of techniques whose understanding by a broad range of users is this paper's focus. To provide context for user-centered communications, including highlighting the types of guarantees private computation provides, this section provides technical background on those techniques. Notions of private computation revolve around two key aspects: \emph{what} is being protected, and from \emph{whom}.
The techniques guarantee particular protections as long as certain assumptions are met. The assumptions can be about
potential adversaries, system complexities, or statistics. When these guarantees are not in place, private information may leak.

A private computation executes a function over an input to produce an output such that there are limits to what can and cannot be inferred by an adversary, even if the adversary possesses some form of additional data. The function enforces the limitations through the use of mathematical protection mechanisms from cryptography (e.g., homomorphic encryption), statistical guarantees (e.g., differential privacy), or a combination of techniques.
Such computations may be between two or more parties, and they may involve trusted third parties. What is being protected within private computation typically falls under one of the following two classes:

\paragraph{Class 1: Private Data Set, Public Results}
Consider a scenario where one or more parties have a (joint) data set and want to release an analysis of the data set.
For example, the Census Bureau may wish to release statistics about the population of a certain region. Abstractly, their analysis $y$ is a function $f$ of the data set $D$, i.e., $y = f(D)$.
The party performing the analysis 
can employ a protection measure like \textbf{differential privacy (DP)}~\cite{dp}, 
which ensures that a single record in the data set $D$ has bounded impact on the analysis $y$. That is, the output distribution of $y$ shifts by at most a factor determined by a privacy parameter specified by the analyst. By bounding the impact of a single record, the individual records in the data set have a measure of protection against being revealed to anyone who accesses the results of the analysis. The analysis becomes a \emph{private} version of the computation. 

The data set $D$ may be distributed among several parties (e.g., $D_1$, $D_2$). 
For example, a government may be interested in the wages of its student population and thus wish to intersect tax filings with universities' registration records. 
Here, the analysis $y$ may be computed as a \textbf{secure multi-party computation (MPC)}~\cite{gmw,yao}, which is a cryptographic protocol enabling the parties to compute the function $y = f(D_1, D_2, \ldots)$ while ensuring that no party $i$ learns anything except $y$ and $D_i$ using techniques like homomorphic encryption. 

\paragraph{Class 2: Private Data Set, Public Subset}
While the previous computations protected all individual data records while revealing the output of a computation, we now discuss approaches that instead aim to publicly (or selectively) reveal a subset of the data. 
Consider a case where parties want to learn additional information about their data or information about a relationship between data sets they each hold individually. 
For example, assume Google holds a set of digital ad views and Mastercard holds a set of credit card transactions~\cite{bergen2018google}.
Google may want to learn which ad views led to credit card transactions, while Mastercard may want to learn which transactions were preceded by an online ad. 
Abstractly, given a common identifier in the data, the two parties could learn the intersection of their sets.
The process of learning this intersection while protecting the respective data sets is known as \textbf{private set intersection (PSI)}~\cite{psi}.
Using PSI, two or more parties can compute the intersection of their data without revealing data they possess outside of the intersection.
Notably, PSI reveals no information about identifiers not in the other party's set, but fully reveals each identifier in common. 
Differential privacy can be used on the data sets for additional privacy~\cite{groce2019cheaper}, and extended forms of PSI can compute a function over the intersection~\cite{pinkas2018Euro}.

\paragraph{Attacks on Private Computation}
So far, we have defined what private computation protects. However, given that some information is revealed intentionally as part of a private computation, there are some risks.  
Recall that we reveal an analysis $y$ as a function of a data set $D$: $y = f(D)$.
Given $y$, it is possible for an adversary to compute the inverse of function $f$ and obtain a set of possible data sets $D$.
This inverse can be computed when given only $y$, but the adversary may also have background knowledge in the form of a probability distribution over the possible data sets $D$, further restricting possible inputs and thus improving the attack.

Inference attacks, a subject of ongoing research, may pose significant privacy risks for subjects in the data set $D$.
For statistical data sets, \textbf{de-anonymization attacks} or other information leakage can come via the execution of summation queries~\cite{malvestuto2006sumqueries}. 

In the case of machine learning, attacks may use queries to the model and other attributes. 
We give a few examples from machine learning where the output $y$ (given to the adversary) is a publicly released machine learning model (e.g., a neural network), the outputs produced by a distributed learning process like federated learning~\cite{fl}, or both.
A \textbf{model inversion attack}~\cite{modelinversion,flinversion} computes the most likely input for one class of the model.
For example, for a face recognition model this can be a picture of the recognized person.
A \textbf{property inference attack}~\cite{propertyinference} computes a property of the records in the data set given a description of the property.
For example, for a face recognition model this can be the ethnicity of the recognized person.
A \textbf{membership inference attack}~\cite{membershipinference,dpbound} computes 
whether or not a given candidate was part of the data set $D$. 
For example, for a medical classification model, this can be whether or not a patient's record was included in the study.

Inference attacks are still feasible if the adversary cannot enumerate all possible data sets $D$ because they only need to estimate the most likely inference.
Differentially private protection mechanisms complicate inference attacks~\cite{dpbound}, but their theoretical analysis is complicated and error-prone~\cite{ours}.

 \section{Related Work}\label{sect:relwork}







\paragraph{Communicating Differential Privacy and MPC}
As detailed in Section~\ref{sect:background}, private computation efforts use a technical mechanism to compute revealed outputs from protected inputs. 
The technical privacy mechanism of differential privacy, and its implications for end users, has received significant attention from the HCI community. 

Researchers have aimed to explain differential privacy using a variety of techniques~\cite{Karegar2022,Cummings2021,nanayakkara2022visualizing,nanayakkara2023,CORMAN2022104859,franzen2022private} and to evaluate whether differential privacy improves users' willingness to share their data~\cite{xiong2020towards,kuhtreiber2021usable,kuhtreiber2022replication,Bullek2017}.
Those efforts include attempts to convey risk using visuals, risk notifications, and metaphors. 
While past work has done an excellent job at investigating ways to communicate about differential privacy, these techniques are too narrow to apply to most other types of private computation. 
Differential privacy provides guarantees of the form ``two neighboring data sets are indistinguishable within some probability,'' and understanding that guarantee requires first understanding the notion of neighboring data sets (i.e., those differing in one row). 
Private computation more generally does not focus on neighboring data sets.
Furthermore, differential privacy's main privacy guarantees result from perturbing, or ``adding noise'' to, a data set. Whereas the aforementioned prior work on differential privacy aims to give non-technical users an intuition around ``adding noise,'' the underlying mathematics of other types of private computation lack an intuitive analogue. 

However, the guarantees and benefits of the other types of private computation we study are arguably more straightforward than differential privacy's guarantees related to neighboring data sets. 
All forms of differential privacy provide a \emph{statistical privacy guarantee}. As further described in Section~\ref{sect:inferenceattacks}, our participants raised concerns about such statistical protections; they felt that privacy guarantees should hold consistently. Other guarantees, such as the information theoretic or computational ones provided by other technical mechanisms, may be viewed more positively by the public and thus should be explained clearly to non-technical users. 

Explanations and potential regulations must also take into account all relevant stakeholders. The limited prior work on private computation mechanisms other than differential privacy has focused on stakeholders other than the data subjects. For example, Agrawal et al.\ investigated the perspectives of specialists like industry professionals, researchers, designers, and policy makers~\cite{agrawal2021exploring}. They found that these specialist participants described private computation as a tool for enabling organizations to overcome `legal gridlocks' related to data sharing. While these specialists acknowledged the importance of end users (data subjects), few prioritized end users' understanding of private computation, increasing the risk that private computation could be used for privacy theater~\cite{smart2022understanding}. Similarly, Qin et al.\ examined the usability of multi-party computation in terms of functionality~\cite{qin2019}, rather than through our lenses of users' perceptions and understanding.

\paragraph{Communicating Encryption} Whereas private computation uses advanced mathematics to compute a function while keeping the function inputs private, encryption uses advanced mathematics to encode data in a way that keeps it confidential. Researchers have studied users' mental models of encryption. For example, via 19 interviews, Wu and Zappala found that users often conceptualize encryption as ``restrictive access control''~\cite{wu2018tree}. Focusing on end-to-end encryption, Abu-Salma et al.\ found that surveyed users lacked confidence in their understanding of encryption and mistakenly believed that others could access information sent using end-to-end encryption~\cite{abu2018exploring}. Subsequent work has aimed to support users' mental models of encryption via improved descriptions~\cite{akgul2021secure,bai2020improving,akgul2021evaluating,distler2020making} or visualizations~\cite{stransky2021limited}. Further, due to gaps in their mental models, users often misunderstand the purpose of authentication ceremonies that help guarantee the security of end-to-end encryption~\cite{herzberg2020secure,fassl2021exploring,vaziripour2018action}.

\paragraph{Privacy Perceptions and Preferences}
Previous work has frequently found users to be averse to their data being shared or sold~\cite{mcdonald2010americans,rader2014awareness,fiesler2018we,shklovski2014leakiness,mayer2021now,kang2015my}. 
Private computation has the potential to counteract this aversion if its guarantees are communicated successfully. 
As a result, it is necessary to study users' awareness, understanding, and motivations of technical tools, including their implications for individual and societal privacy~\cite{story2021awareness,oates2018turtles,atwater2015leading,renaud2014doesn,Cummings2021}. 
Information about individuals may be collected by employers, government entities, and friends. Which collector originally receives the information is one component of the `context,' or social domain, in which information is shared. 
Recent work from Kacsmar et al.~\cite{kacsmarTilMazKer2022} found that different contexts, represented by the number and type of participating companies, have an observable influence on users' perceptions of data-sharing practices. 
Once information is moved to a different context, whether via use or disclosure, it can no longer be assumed to meet privacy expectations~\cite{nissenbaum2019contextual,hanson2020taking}.
Private computation involves two or more organizations contributing their data. That is, private computation inherently results in a change of context that can influence data subjects' perceptions and preferences.










\paragraph{Law and Policy}
To the extent that law formalizes societal norms for enforcement, it is necessary to understand those norms.
Legal notions of privacy are primarily framed in terms of individual protections from government and from corporations, with legal and financial penalties for non-compliance. 
The legal guarantees a company makes are typically communicated within complex privacy policies~\cite{cranor2012necessary,norton2016non,rothchild2017against}.
These guarantees are enforced, as much as they are, by local data privacy laws. For example, 
Canada has PIPEDA, the Personal Information Protection and Electronic Documents Act~\cite{pipedaBrief}. The United States has, among other laws, the Children's Online Privacy Protection Rule (COPPA)~\cite{coppa}, the Health Insurance Portability and Accountability Act (HIPAA)~\cite{act1996health}, and the California Consumer Privacy Act (CCPA)~\cite{CCPA}. Member states of the European Union have the General Data Protection Regulation (GDPR)~\cite{voigt2017eu}. 
Regulations may impact individuals' perceptions and thus necessitate recruiting participants from the same locale. 

Designers of private computation protocols have suggested that these protocols can help ``simplify the legal issues of information sharing''~\cite{pinkas2018Mot} and resolve privacy issues in various domains~\cite{cristofaro2010linear, phasingPInkas2015, kissner2005privacy}. 
However, it takes time to change laws, whereas new technologies are in constant development. Thus, laws may not encompass current and future uses of private computation~\cite{mahoney2020california,o2021clear}. 
Furthermore, our results demonstrate that private computation alone does not resolve privacy issues. Instead, it is critical for consent to be properly acquired, among other aspects of respecting users. 


\section{Methods}


As there has not been much prior work on users' understanding of, and expectations for, the broad range of private computation methods we consider, we employ semi-structured interviews to allow us to follow up on participants' responses and allow participants to ask for clarification. 
All participants received the same set of questions with the order shuffled as appropriate. Appendix~\ref{app:interviewGuide} contains the interview guide. 
We refined our procedure through pilot studies with five participants. Questions that participants found confusing were either removed or clarified. 
We do not include responses from the pilot study in our results. 
The lead institution, located in a country without IRBs, has an institution-level Office of Research Ethics that approved our human-subjects study in an IRB-equivalent process. Ethics Board approval covered the design of the study, consent process, data analysis, and protection of the data collected. Only the researchers at the lead institution engaged in human-subjects research (specific study design, consent process, any interaction with human subjects). Furthermore, only the researchers at the lead institution had access to the data collected.

\subsection{Procedure}\label{sect:procedure}

We developed an interview protocol that addressed the research questions listed in Section~\ref{sec:intro}. We designed our interview questions to gauge participants' understanding and perceptions of key applications of private computation. We include a range of data leakage scenarios to understand how participants perceive risks.


Before starting, we reminded participants that participation was voluntary, that audio was being recorded, and that they were encouraged to ask questions throughout.
The interview proceeded through the parts detailed in the rest of this section: 

\paragraph{Expectations and Term Awareness} The interview began with baseline questions to establish participants' existing perceptions. Participants were asked to ``list some of the ways that you expect companies use data about you and others'' and whether they had ever ``come across'' eight terms related to private computation that we presented in randomized order: ``private computation,'' ``encryption,'' ``hashing,'' ``multi-party computation,'' ``differential privacy,'' ``federated learning,'' ``private machine learning,'' and ``secure computation.'' Terms with which participants were familiar resulted in follow-up questions about where they had come across the term, what they thought its purpose was for companies and individuals, and a request to define the term in their own words.

\paragraph{Private Computation Definition} We then clarified ``private computation'' for participants by defining and comparing a non-private computation with a private computation. After participants had the opportunity to ask questions, they were asked to consider what they thought could be an example of ``a computation where the result could be made public, but the inputs used to determine that result were sensitive and needed to stay private.'' 
 
 \paragraph{Computation Scenario Perceptions}
As one of the key parts of our investigation, we gathered participants' perceptions of, and expectations for, private computation through discussing four scenarios in randomized order. 
Over the course of an interview, these scenarios create what is essentially the process of self-explanation for learning~\cite{angelo2012classroom,chi1989self,chiu2014supporting}. Self-explanation helps learners adjust their understanding of a topic through examples and explaining concepts back to others. Essentially, it is an inductive, generative process of learning private computation rather than a prescriptive process. 

We presented participants with a selection of scenarios in which private computation could be suitably applied. Each scenario consisted of an overall description of the goal of the computation, as well as two ways this goal could be achieved. One way used a straightforward approach involving non-private computation as a baseline. We then presented an alternative approach that employed private computation, enabling participants to compare the two versions. 
For each scenario, we asked participants how acceptable they found each approach, as well as why. Their explanations and reasoning helped us identify what factors most influence perceptions of (non-)private computation. 
We also asked participants what differences they perceived between the straightforward computation and private computation in that scenario, how feasible they considered the private computation to be, and how the company performing data analysis might explain the private computation to users.

In terms of scenario selection, our goal was for each instance to reflect a known real-world deployment, encompass either a conventional "social good" goal or "profit-based" goal, be user-facing, and be something for which there existed clear non-private versions of the computation participants would likely have encountered previously.
We chose our four scenarios---census data~\cite{abowd2018us}, wage equity~\cite{bostonwage}, contact discovery~\cite{de2013private}, and ad conversion~\cite{bergen2018google}---in consultation with our team's cryptography experts based on their impression of the likelihood that private computation would actually be deployed in those scenarios in the real world based on cryptographic feasibility and privacy constraints. These four scenarios encompass three different private computation mechanisms. Both ad conversion and contact discovery are settings where PSI can be deployed. Wage equity efforts can use MPC. Census data can use privacy preserving query procedures. 

In more detail, the \textbf{wage equity scenario} described an organization collecting salary data with the goal of generating a report on inequities. The \textbf{ad conversion scenario} described a credit card company and an online company comparing their data with the goal of determining if digital ads lead to sales in physical stores. The \textbf{contact discovery scenario} described a social media company with the goal of determining whether a new user had contacts that already use the app. Finally, the \textbf{census scenario} described a government body collecting a range of data with the goal of informing policies and resource management, as well as making results public. 
See Appendix~\ref{app:interviewGuide} for the full descriptions and interview guide.


 \paragraph{Inference Attack Perceptions}
We then presented participants with four descriptions corresponding to types of inference attacks.
For each, we gave participants a series of examples of what specifically the company could learn, asking the participant to explain how acceptable they found that situation. For instance, in the case of a membership inference attack, we said, 
     ``One of the participating companies will additionally \emph{be able to learn which specific records in the computed result correspond to you}.''
 The membership inference attack examples included the data set consisting of a set of members of a dating app, a set of frequent drug users, a set of low-income households, and a set of people with a specific health condition. For each example, participants were asked how acceptable it would be if the organizations involved could determine they were a member of the example data set, as well as to explain their reasoning. The other attacks corresponded to model inversion attacks, statistical inference attacks, and property inference attacks.


 \paragraph{General Perceptions}
At this point, participants had engaged with four private computation scenarios, as well as four types of inference attacks. 
To unite these ideas, we asked how the participants thought companies should be communicating to end users how they used data (with and without private computation), as well as what the companies' responsibilities to their data subjects were. 

 \paragraph{Collective Activity}
We concluded the interview with a collective (or connective) drawing exercise that built upon all topics participants engaged with throughout the study~\cite{zamenopoulos2018co,simonsen2013routledge}. We asked participants to pretend they were working at an organization that hoped to use private computation and then consider how they would choose to explain private computation to their customers or clients. Participants were able to write, draw, verbally respond, or use whatever other means of communication they preferred. 
After providing their own explanation, 
participants were shown all previous participants' responses to the question and asked what they would add to that explanation and what (if anything) they would remove from it until they arrived at their final version of the explanation. 
We note that this collective approach integrates input from a range of participants without requiring synchronized timing or a shared location. However, a participant's potential contribution differs based on when they participated, so participants' responses and contributions should not be compared with each other. 
  
\begin{table}[!t]
\centering
\footnotesize
\caption{Participants' demographics, including age range, gender, and highest education completed. Participants indicated whether they have an education or work experience in a tech-related field, as well as in cryptography in particular.\label{tbl:participants}}
\vspace{-1.25em}
\begin{tabular}{ccllcc}
\toprule
\textbf{ID} & \textbf{Age} & \textbf{Gender}  & \textbf{Education} & \textbf{Tech} & \textbf{Crypto} \\
\hline
1            &  18-24   &   Woman      & High School  &         &         \\
2            &  18-24        &  Woman         &  Bachelors        &               &    \\
3            &  35-44        &  Woman        &   High School     &               &         \\
4        &    45-54      &    Man       & Bachelors          &     &          \\
5            &   25-34       &  Man            &  Grad School   &  \cmark       &          \\
6            &   55-64      &   Woman          & Grad School      &           &     \\
7            &   18-24      &  Man           &    Some college      &   \cmark       &           \\
8            &   25-34       &  Woman          &    Bachelors       &            &  \\
 9            &  25-34   &    Man          &    Bachelors      &        &       \\
10           &  25-34 &  Man        &     Grad School      &   \cmark  &  \cmark    \\
11            &   45-54  & Man          &   High School   &       &          \\
12            &  18-24  &   Man      &  Some college    &         &  \\
13            & 35-44 & Woman   &    Bachelors       &          &   \\
14            &  25-34  &   Man          &  Some college   &     \cmark       &      \\
15            & 35-44 &    Man          &  Some college   &         &  \\
16           & 35-44 &    Man          &    Bachelors       &       &  \\
17            & 25-34 &  Man         & Bachelors   &  \cmark       &      \\
18            & 35-44 &  Man          & Grad School           &       &  \\
19            & 35-44  &   Woman        & Some college &         &  \\
20            & 55-64 &   Woman    &   Grad School    &          &    \\
21            & 25-34 &   Woman       &  Some college  &  \cmark       &   \\
22            & 25-34 &  Woman       &  Bachelors         &        &         \\
\bottomrule
\end{tabular}
\end{table}

\subsection{Participant Recruitment}
We recruited participants based in the USA via the Prolific crowdsourcing service using a survey that included demographic information and when they could be available for a synchronous hour-long interview over a video call. We kept interviewing new participants until reaching saturation (no longer finding new themes). We seemed to have reached saturation with just under 20 interviews, but we performed a few extra to be sure. Participants received \$1.45 USD via Prolific for the initial scheduling survey (average time 4 minutes) and an additional \$30 USD for participating in the interview. While most interviews lasted between 50 and 60 minutes, the shortest was 40 minutes and the longest 90 minutes. These times include debugging technical issues (e.g., fixing a microphone).

\subsection{Participant Distribution}
As detailed in Table~\ref{tbl:participants}, we interviewed 22 participants falling in the following age ranges: 18-24 (4 participants), 25-34 (8), 35-44 (6), 45-54 (2), and 55-64 (2). Among participants, 10 identified as a woman and 12 as a man, with no other gender identities being used.
Participants reported working in a variety of fields, including politics, libraries, environmental organizations, education, insurance, health, music engineering, technology, personal assistance, chiropractics, and marketing. 
Participants' highest level of education completed included a graduate degree (5 participants), a bachelor's or associate's degree (8), some college without a degree (6), and high school (3). Further, six participants reported they ``had an education in, or work in, the field of computer science, computer engineering, or IT.'' One of those participants also reported that they ``had an education in, or work in, the field of cryptography.'' We note that the only restrictions on participation was age (18-65) and country of residence. The upper bound was due to requirements our Office of Research Ethics sets for including older participants. We chose not to exclude  the participant who reported cryptography experience as during the interview it became clear their familiarity was overstated. Their responses did not differ from those from participants without that reported background.

\subsection{Data Analysis}
We audio-recorded each interview. We automatically transcribed the audio via speech-to-text software. Afterwards, a member of the research team listened to each recording and corrected the automated transcriptions, as well as grouping responses by question. 

We analyzed this qualitative data using an inductive approach, allowing themes to emerge. Two members of the research team extracted participant responses and then collaboratively clustered them according to similar sentiments and themes using the affinity mapping procedure~\cite{holtzblatt1997contextual,kawakita1991original,scupin1997kj}. Affinity mapping allowed us to employ a team-based, collaborative approach to iteratively identify all aspects participants articulated when discussing their understanding of private computation, as well as private computation's implications. 
As part of the iterative affinity mapping process, after the two researchers formed initial clusters of participant quotes, they reviewed each quote within a theme to see what they had in common and discuss whether the quotes contained any points not encapsulated by others within that theme. Through iteration, we ensured that unique insights were not overshadowed by more prevalent ones.  This process enabled us to capture the full range of attributes participants considered, as well as those that most commonly influenced their opinions.

For example, among responses to the acceptability of the ad conversion case, we identified the following themes: consent, privacy, benefits to the company, and low (perceived) sensitivity. Responses to contact discovery brought out themes of consent as well as benefits, limitations, perceived risk, and data minimization preferences. We reviewed emergent themes with respect to commonalities and differences across scenarios and questions to better understand participants' priorities and concerns. These clusters correspond to the structure of the findings we report in Section~\ref{sec:results}.

\subsection{Limitations}
While we strived to ensure a diverse sample in many aspects, our participants represent a convenience sample and skew young (less than 20\% were age 45+) and educated (69\% had completed a bachelor's or graduate degree).
Our participants are WEIRD (western, educated, industrialized, rich, and democratic), and we make no claims as to our results being representative of other population groups~\cite{schulz2018origins}. 
All of our scenarios are based upon typical cases in North America, where our participants live, and some examples may not be permitted by laws in other countries. Similarly, our scenarios may not cover data analysis tasks that might be both legal and common outside North America.
Finally, as with other response-based studies, we acknowledge the potential for bias towards what participants perceive as socially desirable~\cite{redmiles2018asking}.


\section{Results}\label{sec:results}



We present our results centered on our research questions. In terms of comprehension (RQ1), we present the development of participants' understanding of private computation from their first descriptions through the final explanation they constructed. In terms of perceptions and influence on acceptability (RQ2 to RQ4), we evaluate any changes in perception between scenarios and the reasons participants reported for these changes. This approach enables us to compare the influence of phrasing versus the actual impact as the interview format allowed participants to frame their reasoning in their own words. Thus, we identify themes participants use in their decision-making process when considering data-sharing scenarios, describe how descriptions of private computation influence participants' perceptions of scenarios, and describe the impact private computation has on expectations for companies' responsibilities.

\subsection{Initial Knowledge and Expectations}
Participants' initial expectations for data usage could influence their perceptions of private computation. Thus, 
we present an overview of participants' incoming knowledge and expectations. 

\paragraph{Expectations}
Participants had expectations in terms of what data companies use (purchase history, demographics, search history, salary data, and user preferences), the purposes for which companies use the data (financial gain, improving services, forging social connections, and personalization), and companies' responsibilities with respect to the data (anonymization, preventing re-identification). 
P8 emphasizes that despite being aware of companies' practices, they do not necessarily approve of then:
\begin{quote}
``Even though I don't love that, I expect them to use it like for their marketing purposes\dots grow the bottom line of their business, to make money off of my data, and who I am as a person'' (P8).
\end{quote}
Participants have an expectation that companies are protecting the data entrusted to them, but P18 expressed concern that data usage practices may go beyond what they expect: \textit{``Of course, they may use it for other reasons which I'm not even aware of''} (P18).

\paragraph{Relevant Preexisting Knowledge}
As a proxy for identifying any preconceived notions participants may have about private computation, 
we showed participants a set of relevant terms (see Section~\ref{sect:procedure}). 
There was only one term for which all participants expressed familiarity: encryption. The only other term with any amount of recognition was hashing. However, hashing familiarity was limited to being a data-mapping strategy and not related to cryptographic hash functions. 
All other terms either had no participants reporting familiarity or participants being unable to  place the origins of their familiarity beyond thinking they may have heard the phrase before. In these cases, the participants guessed they either came across the phrase in terms and conditions or in news articles. Thus, we limit detailing previous knowledge to the term encryption. 
 
\paragraph{Source of Awareness}
We surmise that the term encryption is thoroughly embedded in various facets of day-to-day life. Participants responded that they learned of encryption via leisure, education, employment, and when managing finances. However, encryption was not viewed as being particularly relevant to participants' lives:
\begin{quote}
``[It's] something that's used by techie people or politicians or people who are doing nefarious things. I don't think of encryption as guaranteeing things for individuals, like the lay public like myself'' (P6).
\end{quote}

\paragraph{Guarantees}
On one side, participants expressed skepticism as to what tangible protections encryption can provide. Emphasis was made that there are \textit{``no guarantees''} (P16) and that, while it may provide some protections, encryption does not make it impossible for malicious actors to access things. 
For participants that were more optimistic of the protections, encryption was viewed as a means of making it difficult for unauthorized people to access data.

\paragraph{Companies' Purpose}
Some participants responded that encryption is used to provide the \textit{``illusion of security''} (P8), while others thought encryption is used to provide \textit{``customers safety with their data''} (P21). 
Ultimately, whether they had confidence in the protections or not, participants reported that companies use encryption for protecting customer data, protecting proprietary information, gaining customers' trust, or avoiding legal penalties. 

\paragraph{Defining Encryption}
In general, participants' definitions of encryption were not fully comprehensive, but they did show an understanding of encryption at a conceptual level. 
Essentially, participants highlighted that encryption modifies the information to which it is applied. 
These changes were referred to as \textit{``scrambling'' }(P20) and \textit{``masking or disguising'' } (P15) the information.
Further, participants reported that these changes have the goal of providing security to the information so that it cannot be read by unintended recipients. These responses regarding transformations echo what past work termed an ``iterative'' mental model of encryption~\cite{wu2018tree}.

\begin{figure*}[!tb]
    \begin{minipage}[t]{.30\textwidth}
          \includegraphics[trim={0.0cm 0.0cm 0.0cm 0.0cm},clip,width=\textwidth]{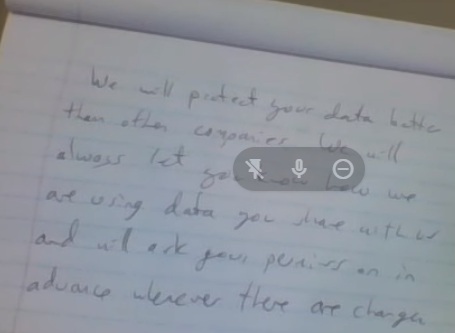}
    \end{minipage}
    \hfill 
    \begin{minipage}[t]{.321\textwidth}
          \includegraphics[trim={0.0cm 0.0cm 0.0cm 0.0cm},clip,width=\textwidth]{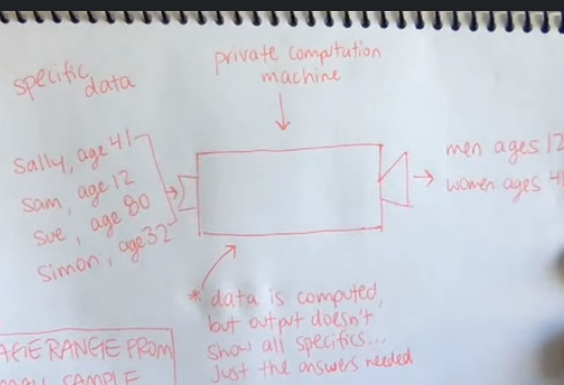}
    \end{minipage}
    \hfill
    \begin{minipage}[t]{.34\textwidth}
          \includegraphics[trim={0.0cm 0.0cm 0.0cm 0.0cm},clip,width=\textwidth]{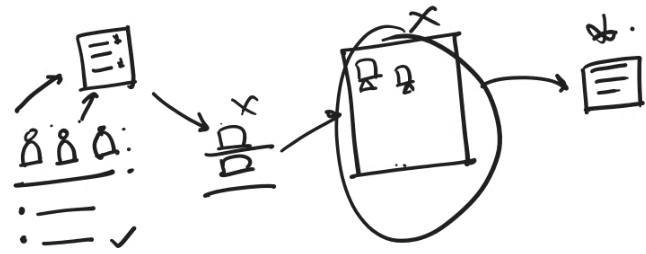}
    \end{minipage}
\vspace{-1em}
\caption{\label{fig:draws} Participants used a range of mediums to convey private computation. Responses included written or typed text, drawn images (digital and paper), and verbal definitions. The above illustrations are from P6, P8, and P10, respectively.}
\end{figure*}

\subsection{Comprehension of Private Computation}\label{subsect:learn}
We asked participants to define the term ``private computation'' at three points throughout the interview as a low-level assessment technique for evaluating learning and understanding of concepts~\cite{angelo2012classroom,chi1989self,chiu2014supporting}. 
We observe an increase in understanding via participants' own explanations of private computation comparing their original response at the start of the interview to their final definition at the end. Over the course of the interview, participants responded to the four different example scenarios, impacting their understanding.
 
\paragraph{First Attempts}
We first asked participants to define private computation in their own words at the beginning of the interview. Specifically, participants were shown an abstract definition and asked to think of an example that could fit the definition. This definition occurred before participants were shown any of the scenarios included in the study. 
Participants struggled to provide an initial definition of private computation. In fact, many participants were unable to come up with any definition. Of those that did provide a definition, they were generally brief and overlapped heavily with the initial definition we had provided. 

In contrast, participants did come up with several examples in response to our prompt for ``an example of a computation where the result can be made public, but the numbers used to determine the result are sensitive and need to stay private.'' 
That said, not all participants came up with an example, some came up with more than one, and some participants changed their mind about their example. See Table~\ref{tbl:examples} in Appendix~\ref{app:table} for the list of examples participants provided. 
The subject domains of the examples included salaries, research studies, and organizations' financial data. 
The public outputs included aggregates, averages, company trends, and post-processed data.
While not all of the examples were appropriate settings for private computation, the participants identified a number of cases that already exist. In particular, participants identified examples that corresponded to two of the scenarios we used later in the study: census data and wage equity.

\paragraph{Second Attempts}
Later in the study, we again asked participants to define private computation. At this point, they had seen all four private computation scenarios and the cases corresponding to inference attacks. For the second explanation, we informed participants that they could use any medium, including drawing a picture, verbal explanations, and writing. 
Participants' second attempt was overwhelmingly more successful than their first. Every participant provided a definition, though with their chosen medium varying; see Figure~\ref{fig:draws} for a selection of responses. Each definition was reasonably accurate, even if not completely comprehensive. Participants included in their descriptions what is being learned and what is being protected as important. Other aspects they suggested to include were how it will benefit the client and what the computation actually is.  In addition to their explanation, participants noted attributes they considered critical to quality explanations. 
Participants particularly emphasized transparency and honesty. Participants also recommended including examples (especially as figures), summaries, and placing visual emphasis on critical points.


\begin{figure}[!tb]
 \vspace{0.5em}
 \rule{8.45cm}{0.2ex}
\emph{Secure computation is a way that a company analyzes your data. The final analysis will be made public at [access location]. However, your specific data is protected and cannot be traced back to you, nor can your specific data points be traced back to you. The analysis will be specifically [example], and this is being done because [purpose].}
      \includegraphics[scale=.35,trim=-1cm 0cm 0cm -1cm]{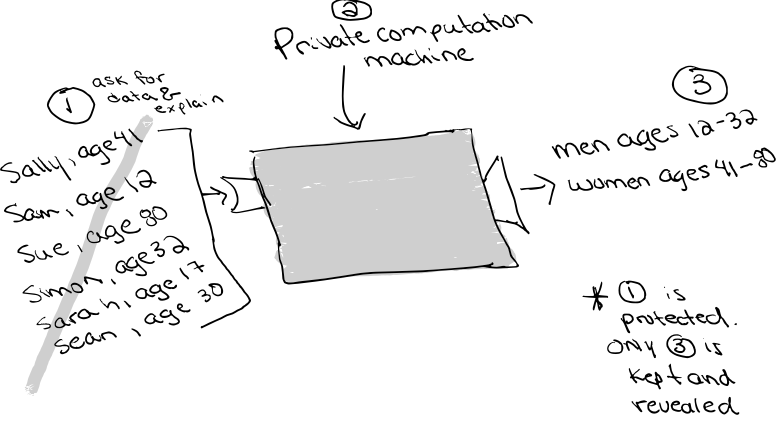}
 ~\\
 \emph{This is the information we're getting from you, but, rest assured, only Part Three will be shown. You can trust us to keep your information private. <If true>This information will only be used for this project and nothing else in the future.}\vspace{-0.5em}
  \rule{8.45cm}{0.2ex}
      \vspace{-2.2em}
      \caption{\label{fig:finalAns} Final explanation of private computation derived from all participants via collective (connective) drawing.}
\end{figure}

\paragraph{Final Explanation}
Participants' final definition is the one they derived after seeing the collective answer from previous participants. In other words, each participant was shown the explanation derived by consensus by the previous participants. They were then asked what they would add or remove to the current explanation with consideration to their own initial response to the prompt. 
Earlier in the study, participants made more dramatic changes, and they often incorporated large portions of their own explanation with smaller components of the current collective explanation. 
As the study progressed, participants made fewer and smaller changes, adding finesse as they identified attributes they considered valuable for an explanation being directed at the public. 

When they made changes to the derived explanation, participants expressed the importance of clarity, accuracy, and conciseness.
Participants emphasized the value of being concise, but that it needs to be balanced with accuracy. For example, P17 noted that the original example would actually not protect the inputs:
\begin{quote}
    ``The only thing I noticed is like, in this example, it's obvious the data is too small, that you can tell like the ages of specific men and women just because there's only two men and two women'' (P17).
\end{quote}

Ultimately, participants made changes to improve clarity across steps in the illustration (consent, input, storage, output) and to emphasize the purpose of private computation. 
For example, P6 found the term ``privacy'' failed to encapsulate what is being done and instead suggested using the term ``secure computation.'' 
They expressed concern that there is a dichotomy between privacy and using customer data such that private computation could never really represent what is being done: 
    \textit{``If you're using my data, then there's no privacy\dots if there's privacy, then you're not using my data''} (P6).
This phrasing choice, which took private computation to secure computation was never reverted by later participants. Further, other participants who noticed the explanation started with a different term expressed support for the change and that \textit{``secure sounds better''} (P8). To improve clarity, P8 introduced a visual example to the explanation. This illustration remained a core component of the final explanation, with other participants making small adjustments. Ultimately, though, later participants expressed an appreciation for the visual (P9, P10, P16-P18, P20). 

The final explanation after all 22 participants, shown in Figure~\ref{fig:finalAns}, encompassed attributes participants emphasized throughout the interview process. 
Within the final answer, there is an explanation providing an overview of the concept, an example that walks the reader through the process (including permission to use the data being requested), the purpose of the computation, and a description of the guarantees being claimed.

As they constructed their explanations, participants did not focus on wanting to know the details of the mathematical mechanism used to achieve the guarantees. Participants trusted that the functionality was feasible without the details, leaving no need for complicated metaphors to prove it (see Section~\ref{sect:initialperceptions}). 
This decision focuses communication of private computation on aspects that are relevant, actionable, and understandable to the populace~\cite{schaub2017designing}. 

Based on the process of collectively creating the final explanation, participants wanted to know the inputs, the outputs, the guarantees, and most of all the purpose of a computation.  
Notably, the final derived explanation specified what was being done and why, provided an illustrated example, and gave a brief explanation of the implications the computation could have for users. 
These components are aligned with the themes that emerged when participants explained the acceptability of the four private computation scenarios, detailed later in this section. This consistency suggests these attributes are critical for obtaining informed consent to private computation. 
The remainder of this section revisits these components and provides insight into why participants considered them relevant.









\subsection{General Impact of Private Computation}\label{sect:initialperceptions} 

For our second research question, we found the following key points. Private computation may influence data subjects' willingness to share their data. However, this influence is not without limits. Participants expressed confidence in the ability for private computation to provide the guarantees described in the scenarios. In many of the presented scenarios, private computation made participants look more favorably upon data sharing. However, as will be discussed in Section~\ref{subsect:conditions}, private computation is not able to completely overcome factors previous work has found to matter to participants (e.g., purpose and consent).

\paragraph{Feasibility of Private Computation} 
Participants overwhelmingly considered the private computations described in each scenario to be feasible. Not only did participants think the scenarios were possible, but they thought such computations may already be happening (e.g., P12 and P13 commenting on census data).
Participants did express concern, however, that companies may not be truthful about what they do with the information they collect (e.g., P22 commenting on contact discovery) and therefore thought it required some sort of enforcement. 
As one participant emphasised, feasibility was not the critical factor: 
\textit{
     ``...it's you know, whether there are guards in place, it's do we have cops to to make sure that they're going to do what they're supposed to do'' }(P16). 

 

Participants acknowledged that private computations could be more expensive than non-private computations, which was stated in the scenario descriptions where appropriate. When they considered the costs, participants included both the company's perspective and their personal views. While participants noted that companies may incur costs from using such computations (P4 and P11), this was not considered a valid reason not to protect users' privacy. Participants even advocated that companies should spend more money on such projects to ensure that users are safe and secure (P2, P20, and P22).

 
 

 

\paragraph{Initial Perceptions of Scenarios} 


Within our sample, participants generally perceived some scenario goals more positively than others. 
Specifically, the scenarios for wage equity and census data were generally perceived positively, with responses clustering on the acceptable end of the scale and with few respondents considering these goals unacceptable.
The scenarios for ad conversion and contact discovery, however, were viewed less positively. For both, responses clustered on the unacceptable end of the scale. For instance, after they considered the contact discovery description, P14 responded that: \textit{
    ``I want some privacy. I don't need 100\%, but I'd like a little bit at least if that's not asking too much'' (}P14).

\paragraph{Potential to Impact Acceptability} 
For each scenario, participants viewed one description corresponding to a non-private computation and subsequently another description corresponding to a private computation. 
The private computation for both the ad conversion and the contact discovery scenarios saw a positive change in acceptability. Furthermore, wage equity had the most significant improvement with no participants reporting the private computation scenario to be unacceptable.

In the private computation scenarios, the stipulations restricting the amount of data revealed and ensuring that companies cannot use the data for any other purposes were cited as improvements over the non-private analogues:
\textit{
    ``Even less of the data...data that is not relevant at all, they modify it to not make it available and I think that's, that's very thoughtful'' }(P9).
When considering the above attributes participants responded that \textit{``it feels a little bit more protected that way''} (P12), \textit{``aligns a smidge more with my values''} (P8), and \textit{``sounds like another layer of security'' }(P19).

Overall, the descriptions corresponding to a private computation tended to improve participants' perceptions of acceptability: \textit{``They're not, you know, over exploiting what they're getting'' }(P22). 
 The exception with respect to acceptability was the scenario for census data:
\textit{
    ``The second one [describing private computation] is kind of saying the same thing\dots they're trying to make it sound a little bit better''} (P19). 
 However, even for the other scenarios, the improvement was not unconditional.
Participants expressed concern for aspects that private computation does not, or cannot, address. Ultimately learning something is the goal of any private computation, and that is not something that can be changed: \textit{``At the end of the day, they're still like learning specific things about me'' }(P7).

\paragraph{Impact on Acceptability Due to Misconceptions} 
While some participants expressed exceptional insight into the risks and implications of private computation, others felt reassurance from its attributes. Unfortunately, not all of the attributes that gave participants reassurance provide the actual protections participants expected. 
We identified two main concepts that participants found reassuring but are known not to provide the guarantees attributed to them.
The first concept that provides false assurances is aggregation. 
For example, P6 described the protection from aggregation as:
\textit{
    ``When it's aggregated, it's lost. It cannot be disassembled. And private does not communicate that in any way shape, or form to me'' }(P6). 
This confidence in averages and aggregation is unfortunately misplaced as there are a number of ways a malicious party could carefully select queries such that they can learn about an individual~\cite{malvestuto2006sumqueries}. 
The incorrect idea that one can ``blend into the crowd'' via averages and aggregates without risk was also evident in participants' responses to the assorted inference attacks they were shown. That is, participants tended to find property inference attacks more acceptable than attacks that targeted an individual. 

Other concepts that provided false assurances were law, policy, and standardization. 
The assumption that the practices are ``legal'' or ``industry standard'' influenced acceptability. For example, P4 specifically stated that if the practice is not an industry standard, then the acceptability would decrease. 
For example, P16 concluded that if companies disclose such practices in their terms and conditions, it must be legal: \textit{
     ``I don't know in the real world if this is legal to do. I would assume it's legal if it's in their terms'' }(P16).
However, while participants expressed confidence that the law protects against improper data-sharing practices, this belief was not universal. 
Some participants stated that such practices do \textit{``not sound ethical even if it's legal'' }(P11).

\subsection{Bounded Impact of Private Computation}\label{subsect:conditions}
For each scenario, we asked participants how acceptable the scenario was and how companies should explain private computation if they use it. 
Across scenarios, participants expressed a range of conditions that influence the acceptability. These conditions demonstrate limitations for private computation in terms of influencing data subjects' willingness to share their data. 

\paragraph{Motives Matter}
When explaining how acceptable they found a scenario, participants said they considered the goals and intentions of the company (P22) and whether they considered the reasons to be just and fair (P11). Goals that benefited society tended to shift their responses toward acceptability. Goals that corresponded to corporate gain tended to shift their responses toward unacceptability.
The scenarios for census data and wage equity were viewed as benefiting society. Participants called census data \textit{``crucial information gathering'' }(P8). When they viewed the census description, participants were influenced by their trust in the government, the importance of the census for society, and how such data is used:
\textit{
    ``If the government is going to spend money, it may as well be based on some data rather than shooting from the hip'' }(P6).
Similarly, the wage equity description was considered to provide an important societal benefit that prioritized fairness and countered discrimination:  
\begin{quote}
    ``Wage equity should be a goal of a civilized society and companies aren't going to do that on their own, so third party organizations come in to try to ameliorate some of the inequity'' (P13).
\end{quote}



Compared to when the organization's goal was viewed as benefiting society, scenarios where the computation benefited the company were received less positively: 
\textit{
    ``This is based on making more money, they're not considering the actual person involved'' }(P11). 
In particular, the ad conversion scenario was seen as exploitative: 
\textit{
    ``Want to determine whether\dots ads are effective? Well, you're still in business, right?\dots That's enough'' }(P16). 
Some participants expressed that they understood why the company would want to perform such computations to determine if money spent on advertising was used effectively. Participants that expressed such understanding were still divided; while some thought it was fair, others thought companies should determine effectiveness without using additional personal data: 
\textit{
``Companies should have their own analytics\dots to figure out their own conversions'' }(P21).

\paragraph{Regulate the Restrictions}
In the census case, the use of private computation actually increased the number of participants that considered the scenario to be unacceptable or completely unacceptable. Participants expressed concern both about the aspect of ``any query'' being permitted as well as about how query restrictions would be determined. Participants worried that companies would exploit such restrictions such that \textit{``it's more like withholding information''} (P18). As a result, they wanted to know \textit{``who is making the decisions regarding the information that's permitted'' }(P8). 


Participants' views were dependent on who makes the restrictions as well as what is restricted. P16 spoke about the importance of allowing the public to replicate results themselves whenever possible. They supported protecting individuals, but emphasized the importance of balancing protections and transparency:
\textit{
    ``If we're talking strictly numbers I lean towards all information available. There shouldn't be any math problem that is hidden'' }(P16).
This view was shared by other participants who also emphasized that the acceptability of such restrictions is highly conditional: 
\begin{quote}
  ``Depending on what information is permitted, it might be good for somebody to know something that they're not permitting through the system, or it might be bad to let people know something'' (P13).
\end{quote}


Finally, some participants considered both descriptions to provide insufficient protections and desired additional restrictions (P5 and P10). These participants suggested a hybrid version of the descriptions to produce what they considered to be a more privacy-preserving version. Specifically, to address their concerns, they suggested a query variant that only allows aggregate-based (or average-based) queries while also preventing inferences.



 
\paragraph{Divulge the Details}

Identifying what information individuals prioritized in their decision making is key to ensuring that the necessary information is communicated in the future.
Participants mentioned a number of details they indicated as influencing acceptability. 
In particular, participants who responded that a scenario was neutral or unacceptable emphasized that further information was required before the scenario could be considered acceptable. 
First and foremost, participants wanted to know when their data was being used: 
\textit{
    ``That [the data] is being used. What's being done with it. The other company that is involved, that is having access to it. If it's going to be ongoing''} (P17).
Participants also wanted to know specifically how the data is being used. They wanted to know who is performing the computations and why they are being done. They wanted to know for how long the data is kept, how the data is protected (including the limits of those protections), and the implications for their privacy if their data is used in these ways. 

For some participants, a failure to provide details or implement any of the protections the organization claims are reasons to decline to participate in private computation.
In other words, even when private computation is employed, participants care about appropriate flows of information~\cite{nissenbaum2019contextual}. Participants want to be allowed to judge if a flow is appropriate for themselves. To do so, they require details with respect to the information flows. 






 

\paragraph{Consent Above All}
Participants' desire to be informed about information flows would also give them autonomy over their data: 
\begin{quote}
    ``Every time your data is used in some kind of computation, you should be specifically alerted by the company. They shouldn't be able to do private computations\dots without you being aware of it'' (P13).
\end{quote}
A theme that emerged across all scenarios was consent, as well as the importance of choice and communication as part of meaningful consent:
\textit{
    ``If they don't prompt you, then completely unacceptable. If they do prompt you, then completely acceptable'' }(P17). 
Further, P1 and P16 both emphasized that consent is not a one-time thing. Companies need to be informing individuals periodically, or \textit{``every step of the way'' }(P16), about how their data is being used. As part of this process, the company needs to ensure that the data subject continues to consent: 
\textit{
    ``When they sign up for the credit card and periodically, they should be reminded that all of their data is, you know, being sold to other companies'' }(P1).

In cases where participants may want to withdraw consent, the means to do so should be clear and accessible. Companies need to be \textit{``giving simple directions of, you know, where to go to opt out on the application'' }(P4). Such directions support individuals who change their mind about data use, as well as those who did not initially understand or intend to agree:  
\textit{
   ``If a person finds out they signed something they really didn't understand, they can have a way to retract their permissions'' }(P13). 

The final attribute participants emphasized as critical for consent is the use of clear and transparent communication. That is, companies need to be proactive and not just rely on legal contracts to avoid liability.  
For instance, communication about data use should not be buried in terms and conditions nor obfuscated by legalese: 
\begin{quote}
    ``Be more upfront about how they're using our data instead of varying it in like really wordy terms and conditions in language that the average person like myself\dots can't understand very well'' (P1).
\end{quote}









\subsection{Risks for Unique Threat Models}\label{subsect:risksImplications}
In addition to the risks discussed toward the end of the study (e.g., inference attacks), participants highlighted other risks they perceived as possibilities. 
Participants questioned the implications of private computation and identified a number of risks associated with certain deployment contexts. 
Both P13 and P19 mentioned risks associated with the goals of the scenarios, regardless of the use of private computation. 
Individuals can be in situations where computing connections could put someone's safety at risk. For instance, in the contact discovery scenario, P19 expressed concern that such connections could reveal someone's internet presence to an abusive ex or someone for whom they have a restraining order: 
\begin{quote}
    ``[Via] common contacts now he all of a sudden has a friend who has her information and now he has her information. If through the tangled web you could be able to find people\dots that's a growing problem'' (P19).
\end{quote}
Such risks are not necessarily resolved with a technical solution, such as private set intersection, but instead highlight the importance of informing users and respecting their own risk assessments.


\subsection{Inference Attacks and Acceptability}\label{sect:inferenceattacks}
Before we presented any of the inference attacks, one participant independently brought up the concern that organizations might make inferences: \textit{
    ``If you're only giving limited information, you might wonder if they're gonna acquire other personal information about you from that''} (P22). 
Participants also expressed concern that they \textit{``can't really figure out\dots the implication'' }(P6) of computations or \textit{``how it could be exploited'' }(P15). 
The concern is that companies may request limited information, but learn more via other means.

When presented with specific examples of information leakage, two risks associated with inference attacks were most concerning to participants. First, participants worried about any instance where an individual is identified (e.g., membership inference attacks). Second, they worried about any instance where a group of people could be discriminated against (e.g., in certain property inference attacks). 
Across all inference attacks, the perceived sensitivity of the data affected acceptability. 
Location data, health data, sexual orientation, and religion are cases where the type of data was deemed to be especially sensitive. 
Of particular concern was health data. Participants, who were all located in the United States, expressed concern that their insurance company would get this information:
\begin{quote}
    ``If that information then got shared with like my insurance company [they] would then decide to raise my rates because maybe I am at an increased risk for heart disease'' (P1).
\end{quote}
Among participants, there was concern that the inferences made through the attacks could be used in malicious ways and to propagate bias and discrimination: 
\textit{
    ``What this data is going to be used for, the state of it, should be used to to propel humanity forward.  Not hold, not keep people back'' }(P16).
    
With respect to the inference attacks, some participants viewed all such attacks as unacceptable  
 because the companies were \textit{``not supposed to have that information, period'' }(P6). 
However, we  did observe that inferences that target groups rather than individuals were viewed less negatively. Properties of groups were generally perceived to be somewhat more acceptable. However, this trend was conditional upon the specific property and that property's potential implications for individuals and society. For instance, if the property could be used to \textit{``manipulate the populace'' }(P13) or was \textit{``rude''} or \textit{``discriminating''} (P22), participants found it less acceptable.

For conditional attacks, information leaks only occur probabilistically. However, this was not necessarily viewed positively by participants:
\textit{
    ``It's based on what that record is in relation to even if it needs to be protected''} (P16). 
Many found it unacceptable regardless of the percentages and stated that the percentage was irrelevant. The three participants who reported a tipping point placed it at a 50\%, 25\%, or 1-2\% chance the exact record would be learned.

\subsection{Expectations for Responsibilities}\label{subsect:responsibilities} 
While private computation positively impacted participants' perception of the scenarios, these perceptions were impacted to a greater degree by other factors. The absence of attributes like transparency and communication would lead to a more negative reception even when employing private computation:
 \begin{quote}
     ``It takes more effort, though, and time on the companies to do that. But if they're willing to, I think it might add a lot to their, you know, trust in the credentials of that company'' (P22).
 \end{quote}

 Participants identified responsibilities for companies, governments, and even themselves as individuals. They felt companies have the greatest responsibility with respect to the law, protecting user data, and treating data with respect. Governments' responsibility was to protect individuals by creating and enforcing policy.

\paragraph{Proactive and Transparent Communication}
An individual's ability to protect themselves is almost inconsequential without support. For example, after expounding on how a company's priority is financial gain, one participant expressed concern for how data subjects are supposed to learn what they need to have data autonomy:  
\textit{
    ``How do I protect myself and who teaches me how to protect myself? Who's responsible for teaching me how to protect myself?'' }(P6).

When using customer data, companies need to be upfront about their actions, yet also provide greater granularity of control: 
\textit{
      ``It's my responsibility at this point, quite honestly, which is really hard because it's very confusing'' }(P6). 
For example, rather than giving data subjects a vague description, companies can be more specific:
\textit{
    ``It doesn't really give much more information on what type of data is being used''} (P12). 
That is, participants suggested having companies detail what is being protected and what risks persist even when employing a privacy-preserving technique.

\paragraph{Respectful Treatment} Participants expected companies to protect the data entrusted to them using the ``best'' security measures available as that data is not just some abstract input to compute over. In other words, they wanted companies to re-humanize the data entrusted to them as all of the data they hold corresponds to an \textit{``actual individual person with a name, a face'' }(P9). 
Participants expected companies to treat data with respect. 
Reckless treatment of data can have real consequences for people:
\begin{quote}
    ``I think the ultimate responsibility is to use it with caution. To protect people's privacy.
    It's up to the company to make sure they only share to the extent the person allowed them to.'' (P9)
\end{quote}
Respecting the people who are represented by the data requires companies to exercise clear communication. Without transparency into data-sharing practices, data subjects lack autonomy. 



Participants also felt companies need to acquire explicit and ongoing permission for the collection and use of data regardless of the use of private computation. One participant even hypothesized that data-sharing practices would be more positively received if there were less obfuscation and manipulation: 
\textit{
     ``A big social outcry\dots that could really be prevented if they were open from the very beginning. If people just knew, they wouldn't be so spooked by it'' }(P9).

\paragraph{Government Regulation and Enforcement}
Participants also expressed a level of reassurance toward a scenario in which companies comply with government regulations. However, the nature of these regulations was not always clear to them. 
While some participants called for clearer regulations, some  directly called for the practice of companies selling data to be made illegal:
\textit{
    ``They need to stop selling our information in general\dots Passing that information to a company, I just think it should be illegal'' }(P19). 
Most participants felt private computation does not impact companies' legal responsibilities: 
\begin{quote}
    ``Health is a sensitive topic and there are already legal protections for health information and so on\dots I don't see how why this addition of technology should should change those protections'' (P16).
\end{quote}
Participants made suggestions as to how the law can be enforced via independent third parties. For instance, P21 suggested a third party could perform compliance checks and P1 suggested an independent entity could review points critical to consent. 
The third party could also determine the best way to communicate to users about how their data is being used. They could also determine what information users need to make informed choices about their data. 

\section{Discussion}
Across participants, each individual demonstrated increased understanding (via explanatory evaluation) and communicated to the researchers factors related to private computation that influenced their perceptions of these practices. The reasoning expressed by our participants included both traditional aspects for data sharing (purpose and transparency) as well as technical guarantees (statistical-inference protection, property-inference protection, and membership-inference protection). In this section, we 
discuss how to better communicate to data subjects about private computation. 

\paragraph{For Researchers} 
The use of private computation often improved participants' perceptions of the acceptability of data sharing. In other words, participants recognized the value of applying private computation. 
However, these improvements were neither universal nor unconditional. 
Private computation did not resolve participants' concerns in all scenarios. 
We recommend that future work build on the description our participants collectively created (Figure~\ref{fig:finalAns}) while aiming to improve communication about private computation. 

The description participants created focused on the purpose and implications of data sharing, rather than the complex mathematical underpinnings. Notably, we found that participants did not feel they needed to understand how private computation worked mathematically to find it plausible and feasible. Earlier work on encryption similarly found that users trust the mechanism works without necessarily understanding the mathematics, albeit while holding some misconceptions~\cite{wu2018tree,stransky2021limited,distler2020making,distler2019security}. As long as users trust the entities using private computation, our study suggests that private computation can make more types of data sharing acceptable. 

However, private computation's protections both have limits and create trade-offs. For example, for private set intersection, a malicious participating entity could fraudulently add non-members to its own list to determine whether those individuals are in another entity's database. In this sense, the privacy guarantees depend on the honest participation of each entity. Given that our participants closely scrutinized the purpose of data sharing in evaluating acceptability even with private computation, future communications might further highlight the need to trust participating entities.

As mentioned earlier, differential privacy provides probabilistic privacy guarantees, whereas other types of private computation often provide more straightforward guarantees. Future work ought to compare users' perceptions of these types of guarantees more directly, such as whether participants would prefer their data be protected by differential privacy or other types of private computation. Researchers should also evaluate whether and why probabilistic guarantees are appropriate and sufficient for their systems.


\paragraph{For Lawmakers and Policymakers}
Regulations covering private computation should account for how descriptions of such practices influence data subjects' willingness to share their data, potentially more so than the actual guarantees. 
For example, confidence in the protections of aggregated computations and averages may be misplaced~\cite{malvestuto2006sumqueries}. To ensure that dishonorable organizations do not use this confidence to propagate dark patterns~\cite{bosch2016tales}, regulations must require that companies communicate data sharing's implications. It is impossible to express all possible implications that could result from a computation. Nonetheless, laws should require companies to make explicit what types of protections are impossible or unlikely. 

In prior work, some experts felt private computation could help organizations overcome `legal gridlocks' related to sharing data~\cite{agrawal2021exploring}. In contrast, one of our key results is that private computation was not a panacea for participants' concerns. While participants generally preferred the private computation variant we showed over its non-private analogue, their attitudes still depended most heavily on the purpose of data sharing and consent processes. As a result, laws and regulations ought to consider private computation as a best practice for data sharing despite its potentially heavy computational costs, rather than a silver bullet enabling previously unacceptable flows of personal information. 

\paragraph{For Companies} 
Private computation techniques are a powerful tool that can increase trust from their users when used as a data-minimization technique. That is, a company should employ appropriate private computation tools for data analyses that are already part of their workflow. When the company adds new types of data collection or flows, private computation alone is insufficient. 
Communication should be transparent, accessible, and clear. The onus is on the companies to ensure they obtain informed consent. 
While meaningful consent is a challenge to achieve, it goes a long way toward fostering user trust. 
Even when using private computation, companies must communicate with the same level of transparency, including details related to how the computation is used and what the company might learn from the computation.



\section{Conclusion}
While technical solutions are a powerful tool for protecting data, such protections do not directly correspond to personal privacy protections. The data being protected in these scenarios is not just an abstract concept, but instead is a placeholder for individuals with real lives and all the complexities that entails for their threat models. 
Researchers, data collectors, and policy makers need to remember that the protections provided by protocols and constructions do not---and cannot---encompass the full range of risks experienced by individuals in society. 
Technical privacy solutions must be conscious of the space in which they may be deployed. 
As we found in our interview study, technical solutions do add value, but that value must not be overstated. The data on which we compute so abstractly is very concrete for the people whose lives generated it.

\bibliographystyle{ACM-Reference-Format}
\bibliography{bibliography}


\appendix
 
\section{Additional Table}\label{app:table}

\begin{table}[!h]
\centering
\scriptsize
\caption{Participants' responses to ``an example of a computation where the result can be made public, but the numbers used to determine the result are sensitive and need to stay private.'' The table only includes responses about which participants remained consistent during the interview.
\label{tbl:examples}}
\vspace{-1.75em}
\resizebox{\columnwidth}{!}{%
   \begin{tabular}{lll}
\toprule
\textbf{Example data} & \textbf{Private Data} & \textbf{Public Output} \\
\hline
 (P1) Individual income, education completed    &    Individuals' incomes   &  Mean income by education   \\
(P2) Voting &      Individuals' votes             &     Result counts                  \\
(P3) Research study      &     Participants      &  Study data \\
(P5) Voting      &    Individuals' votes  &      Eligible voters  \\
(P6) Income, location   & Households' income   &    Mean income in a region                  \\
(P7) Salaries   &         Individuals' salary              &      Average salary      \\                 
(P9) Financial organizations' data      &  Customer data    &     Financial trends                  \\
(P10) Telescope data              &  Raw data    & Post-processed data         \\                  
(P12) Personal data     & i.e. age, demographics          &  Averages         \\     
(P13)   Netflix views     &  Viewer distributions &  Report on top service         \\  
(P17) Salaries   &         Individuals' salary              &      Average salary       \\     
(P18) Political surveys    &      Individual responses             &   Aggregated conclusions          \\   
(P19) Profits    &    Beneficiaries  &    Donations           \\ 
(P21)  Elections    &   Individuals' responses    &          Poll numbers      \\ 
(P21)  Infection disease studies    &   Collected data    &    Results     \\ 
\bottomrule
\end{tabular}%
}
\end{table}

\newpage

\onecolumn

\section{Interview Guide}\label{app:interviewGuide}

\begin{scriptsize}
\textit{The order of the terms (a-h), the four scenarios (wage equity, census data, ad conversion, contact discovery), the four cases (one to four), and the examples within each case (a to d) were randomized.}

Welcome. Today we are going to be talking about a topic that may be new to you. 
We're currently studying public sentiments and understanding of novel data science techniques. We're interested in learning about what people expect and what questions they want addressed if their data is being used for data science by a company. The interview process helps us to understand these expectations and based on them, to make design recommendations for other researchers and policy makers.
Please let us know at any point if you have questions. 
Before we start, I just want to make sure you have a something to write with/on, pen and paper.
Throughout the interview, we’re going to go through four types of questions, some general, some about terminology, some about types of data sharing, and some about explaining how data is used. 
On average I expect this interview to take $60$ minutes. Do you have any questions or concerns before we start?

To get us started, I'm going to ask you a general question on the topic. For the question, just state as many answers as come to mind and let me know when you're done. Please list some of the ways that you expect companies use data about you and others. 


Next, we are going to talk about approaches to data sharing that focus on 'how' the data is shared.
We are going to go through a series of terms and I'll ask you if you are familiar with them, and some follow up questions: \textit{(a) Private Computation; (b) Encryption; (c) Hashing; (d) Multi-party Computation; (e) Differential Privacy; (f) Federated Learning; (g) Private Machine Learning; (h) Secure Computation}

Have you come across the term \textbf{[(a) through (h)]} before? 
 \begin{enumerate}
     \item  (if yes) Where have you come across the term before? 
     \item (if yes) What kind of guarantees do you think it provides to individuals? Some examples?
     \item (if yes) What do you think the purpose or goal is for a company using this?
    \item Please try to define the term in your own words 
 \end{enumerate}


We're now going to introduce the term private computation. A \textbf{computation} is just a calculation (generally in math). For instance, determining the largest number from a list, determining the average, determining a sum. A \textbf{private computation}, is a computation that tries to limit the information revealed by the result. It attempts to perform a computation (such as an average, sum, max), and share the result without anyone learning the values used to find the result.

\begin{enumerate}
    \item What do you think is an example of a computation where the result can be made public, but the numbers used to determine the result are sensitive and need to stay private?
     Follow up: what is sensitive and what is not in the example.  
    \item How would you describe private computation in your own words?
\end{enumerate}



We are now going to talk about some different ways companies can work with client data.

\noindent \textbf{Scenario 1 (wage equity):} An organization aims to identify salary inequities across demographics. They reach out to individuals and employment organizations about their salary data. The organization conducts an analysis over the salary data and produces a report on salary inequities. The organization acquires the data for the analysis such that...
  How acceptable is the organization's goal?  Scale: (completely unacceptable, unacceptable, neutral, acceptable, completely acceptable)
        \begin{enumerate}
            \item \dots salary data is shared directly. They receive the salary information of individuals from the individuals or employers via a web-based tool.
            \item \dots salary data is submitted in a modified form privately (with technical and legal protections) via a web-based multi-party computation (MPC) tool. The technical protections prevent the identification of individuals' salary input from the final report. It also protects those who contributed their salary information from being connected to the salary information they provided (though does not prevent it from being known that they were a contributor). Using this technique can be more expensive for the analysis  and they cannot use the data for any other purpose. 
        \end{enumerate}

\noindent \textbf{Scenario 2 (census data):} Census data is acquired from citizens of the country by the governing body. It includes information with respect to their age, gender, occupation, income, place of residence. The governing body analyses the data it acquires to inform policies and resource management. It can also make the results of the census available to researchers or the public by...
  How acceptable is the organization's’ goal?  Scale: (completely unacceptable, unacceptable, neutral, acceptable,  completely acceptable)
 
           \begin{enumerate}
            \item allowing aggregate/statistical queries (e.g. averages, sums, etc.) over the original data.
            \item allowing any query, but restricting individuals making queries from performing queries that allow them to make inferences/learn more information than is permitted. This means that some questions cannot be answered by querying the data. 
             \end{enumerate}

\noindent \textbf{Scenario 1 (ad conversion):} An online ad company wants to determine whether ads shown to its users lead to sales in physical stores. They reach out to a credit card company, which has transaction data for physical stores to compute whether there are purchases connected to their ads. The two companies perform the computation such that...
  How acceptable is the organization's’ goal? Scale: (completely unacceptable, unacceptable, neutral, acceptable,  completely acceptable)
             \begin{enumerate}
            \item \dots they each share their data sets. The credit card company shares the purchase data in physical stores and the online company computes the correlation to online identities locations and online ad views. 
            \item \dots the credit card company shares a modified version of their records. The credit card company shares the modified data such that the online company can only identify the financial records that correspond to its users. That is, the information on the other credit card clients (that do not use the online service) is not available to the online company. Using this technique can be more expensive for the company and they cannot use the data for any other purpose. 
        \end{enumerate}
\noindent \textbf{Scenario 4 (contact discovery):} A social media app wants to connect users that are already contacts with one another. The social media app has a list of contact information (its users) and the new user has a list of contact information (their friends etc).
    The app wants to determine the common contacts between the new user and the existing app users (the intersection). Note that not all of the new users contacts may use the social media app and not all users of the app are contacts with the new user. The social media app can connect the new user to existing users by performing a computation such that...
 How acceptable is the organization's goal?  Scale: (completely unacceptable, unacceptable, neutral, acceptable,  completely acceptable)
     
        \begin{enumerate}
            \item \dots the new user shares all their personal contact information with the social media app. 
            \item \dots the new user shares a modified version of their personal contact information. The new user shares the modified data such that the social media company can only identify the new users' contacts that already use the social media app. That is, the other contacts (who do not use the social media app) are not available to the social media app. Using this technique can be more expensive for the company and they cannot use the data for any other purpose. 
        \end{enumerate}

For each of [A], [B], [C], and [D], we asked the following questions: 
\begin{enumerate}
 \item  How acceptable is it if the company uses (a)? Explain.
            (completely unacceptable, unacceptable, neutral, acceptable,  completely acceptable)
                \item  How acceptable is it if the company uses (b)? Explain.
            (completely unacceptable, unacceptable, neutral, acceptable,  completely acceptable)
    \item What differences do you expect there should be (if any) if a company chooses to use \textbf{(b)} instead of \textbf{(a)}\dots 
    \begin{enumerate}
        \item in general?
       \item in terms of how companies inform their clients that their data is being used?
      \item in terms of what companies inform their clients about when their data is being used?
    \end{enumerate}
    
    \item How feasible/possible do you think it is for a company to use \textbf{(b)} instead of \textbf{(a)}
    \item How should a company be explaining the technique \textbf{(b)} to their clients if they use it?
 
\end{enumerate}


\noindent \textbf{Case 1:} One of the participating companies will additionally  \emph{be able to learn which specific records in the computed result correspond to you. How acceptable is it if the records that correspond to you are…}
\begin{enumerate}[a)]
    \item \dots your salary information? Explain.
    \item \dots your credit history (e.g., credit score, mortgage status)? Explain.
    \item \dots your location history (e.g., coordinates corresponding to your home, place of employment, etc.) Explain.
    \item \dots your genetic markers (e.g., for heart disease, cancer, etc.)? Explain.
\end{enumerate} 
 
\noindent \textbf{Case 2:}
One of the participating companies will additionally \emph{be able to learn if records of you were used to perform the computation. How acceptable is it if the records they learn correspond to you are in a data set of\dots }

\begin{enumerate}[a)]
    \item \dots low-income households (and thus learn that you  are in a low income household)? Explain.
    \item \dots dating app members (and thus learn that you use that dating app)? Explain.
    \item \dots people with a specific health condition e.g., diabetic, high-blood pressure, autoimmune diseases (and thus learn that you have that specific health condition)?     Explain.
    \item \dots frequent drug users e.g., alcohol, marijuana, others (and thus learn that you are a frequent user of that drug)? Explain.
\end{enumerate} 

\noindent \textbf{Case 3:}
One of the participating companies will learn \emph{properties for groups}. A group could be people with glasses or any other attribute corresponding to a group of people such as demographics. How acceptable is it if a company can learn, for example...

\begin{enumerate}[a)]
    \item \dots glasses owners prefer shopping online? Explain.
    \item \dots women prefer shopping online? Explain.
    \item \dots glasses owners have poorer spending habits than non-glasses owners? Explain.
    \item \dots women have poorer spending habits than non-women? Explain.
\end{enumerate}

\noindent \textbf{Case 4:} 
When two companies perform the private computation, if one of the
participating companies possesses other additional information (e.g. statistics) \emph{they can infer the exact value of a record  used in the computation}. 
How acceptable is it if a company can always learn whether an exact record was contributed by the other organization? Explain.

\begin{enumerate}[a)]
    \item How acceptable is it if a company can always learn whether an exact record was contributed by the other organization? Explain.
    \item Is it more or less acceptable if a company can accurately learn the record contributed by a different company only 75\% of the time? Explain.
    \item \dots 50\% of the time? Explain.
    \item \dots 25\% of the time? Explain.
    \item To you, at what point (percentage) does this become unacceptable/acceptable? Explain.
\end{enumerate}

\begin{itemize}
      \item How does it impact the acceptability if additional information has to be known to learn the values?
    \item How does the information that needs to be known influence the acceptability?
    \item How does the likelihood the additional information is known influence the acceptability?
\end{itemize}

\begin{enumerate}
    \item In general, how do you think companies should be communicating to their customers/clients about how they use customer/client data in general? 
    \item In general, how do you think companies should be communicating to their customers/clients about how they use customer/client data if they use {private computation} for the process?
    \item In general, what do you think are companies' responsibilities when using your data in these computations? Follow up depending on response: in terms of data protection responsibilities? 
\end{enumerate}
 
 
The last thing we are going to do is an exercise called co-design. 
Even though you may have just learned about these techniques, we want you to think about how you would communicate these techniques to someone. There are no right or wrong answers. Imagine you work for a company that wants to use private computation. How would you communicate these practices to your clients?
You can draw, write, verbally explain, etc.  
\textit{[Show participant the previous suggestion.]} What would you add to or remove from yours based on it? What would you add to or remove from the previous one?

\end{scriptsize}

\end{document}